\documentclass[12pt]{article}
\usepackage{amsmath,amssymb,epsfig}

\usepackage{color}
\input{colordvi.tex}
\def\unit{{\relax{\rm 1\kern-.26em I}}}

\newcommand{\tr}{{\rm Tr}}

\makeatletter

\renewcommand\section{\@startsection {section}{1}{\z@}%
                                   {-3.5ex \@plus -1ex \@minus -.2ex}%
                                   {2.3ex \@plus.2ex}%
                                   {\normalfont\large\bfseries}}

\renewcommand\subsection{\@startsection{subsection}{2}{\z@}%
                                     {-3.25ex\@plus -1ex \@minus -.2ex}%
                                     {1.5ex \@plus .2ex}%
                                     {\normalfont\normalsize\bfseries}}

\makeatother

\begin{document}

\baselineskip=18pt  
\numberwithin{equation}{section}  
\allowdisplaybreaks  



%
%


\thispagestyle{empty}

\vspace*{-2cm}
\begin{flushright}
\end{flushright}

\begin{flushright}
KYUSHU-HET-165
\end{flushright}

\begin{center}

\vspace{1.4cm}

\vspace{1cm}
{\bf\Large Thermal Effects on Decays } \\
\vspace{0.3cm}
{\bf\Large of a Metastable Brane Configuration}
\vspace*{1.3cm}

{\bf
Yuichiro Nakai$^{1}$ and Yutaka Ookouchi$^{2}$} \\
\vspace*{0.5cm}

${ }^{1}${\it Department of Physics, Harvard University, Cambridge, MA 02138, USA}\\
${ }^{2}${\it Faculty of Arts and Science \& Department of Physics, \\ Kyushu University, Fukuoka 819-0395, Japan  }\\

\vspace*{0.5cm}

\end{center}

\vspace{1cm} \centerline{\bf Abstract} \vspace*{0.5cm}

We study thermal effects on a decay process of a false vacuum in type IIA string theory. At finite temperature, the potential of the theory is corrected and also thermally excited modes enhance the decay rate. The false vacuum can accommodate a string-like object.
This cosmic string makes the bubble creation rate much larger and causes an inhomogeneous vacuum decay. We investigate thermal corrections to the DBI action for the bubble/string bound state and discuss a thermally assisted tunneling process. We show that thermally excited states enhance the tunneling rate of the decay process, which makes the life-time of the false vacuum much shorter.

\newpage
\setcounter{page}{1} 



\section{Introduction}

The idea of the string landscape may suggest that there exist a large number of metastable vacua in string theories \cite{landscape1,landscape2}. If this is true, to reveal the early stage of the universe, it is quite important to study the vacuum selection and the life-time of the vacua. We explore (inhomogeneous) decay processes of false vacua in string theories initiated in the paper \cite{KO1}. The authors of Ref~\cite{KO1,KO2,KNO} studied catalytic effects on the vacuum decay induced by solitons. The solitons enhance the bubble creation rate around them and make the life-time of vacua shorter. In this paper, we mainly focus on thermal effects on the inhomogeneous decay process of a false vacuum in Type IIA string theory \cite{KO2}. The authors of Ref~\cite{EHT} studied the existence of the cosmic string which can be interpreted as a vortex generated by spontaneous breaking of a $U(1)$ symmetry in the field theory limit. The idea of the inhomogeneous decay of a false vacuum by solitons was originally discussed in \cite{Pole,Hosotani,Yajnik} in the context of grand unified theories and recently revisited from the viewpoint of phenomenological model building \cite{Kumar,Rstring,Lee}.

The thermal effects on the brane configuration on which we focus here are twofold. One is the thermal potential generated at finite temperature. The brane becomes non-extremal and feels nonzero forces from the other branes in the setup. This effect can be interpreted as the induced thermal potential of the DBI action. The other thermal effect can be described by the thermal DBI action of the $D$ brane in which the (Euclidean) time direction has a periodic boundary condition. At finite temperature, the vacuum decay starts from not only the ground state but also thermally excited states. In general, we have to treat these two types of thermal effects at the same time. However, by taking an appropriate limit, we can study each effect separately which makes the underling physics clearer.

The plan of this paper is as follows. In section 2, we quickly review the type IIA brane configuration which we will consider throughout this paper. We identify the false and true vacua in this brane setup \cite{OoguriOokouchi,Franco,IAS,Giveon,Kutasov}. In section 3, we discuss the thermal potential of the theory. The dominant contribution to the potential comes from the attractive force of the $NS$-five brane, which stabilizes the false vacuum. In section 4, we consider the thermal effect on the DBI action and calculate the decay rate of the vacuum induced by thermally exited states. Section 5 is devoted to conclusions and discussions.

\section{A false vacuum in Type IIA string theory}

In this section, we first introduce our setup. We review the vacuum structure in type IIA string theory discussed in Ref~\cite{OoguriOokouchi,Franco,IAS,Giveon,Kutasov} (see Ref~\cite{KOOreview} for a recent review article). Here, we assume the string coupling is small, $g_s \ll 1$, and neglect gravitational effects. Then, we introduce a cosmic string which can exist in the false vacuum. As we will see later, the cosmic string plays an important role in the vacuum decay. 

\subsection{The $NS$-five/$D4$ brane system}

\begin{figure}[!t]
\begin{center}
 \includegraphics[width=.55\linewidth]{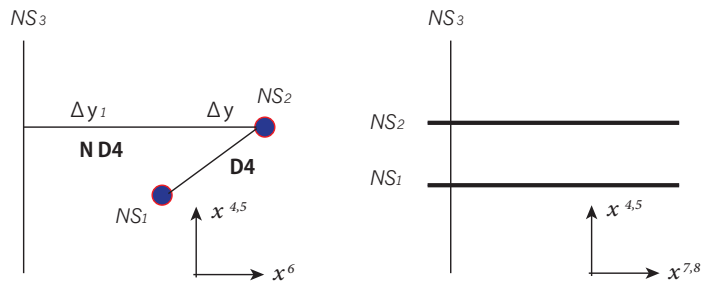}
\vspace{-.1cm}
\caption{The brane configuration without deformation.}
\label{braneFig}
\end{center}
\end{figure}

 Consider three $NS$-five branes extending in the internal space of string theory. The $NS_1$ and $NS_2$ branes are extending in the $x^{7,8}$ directions and are placed at $(x^{4,5,9}, x^{6})=(0,y_1)$, $(x^{4,5,9}, x^{6})=(\Delta_4 ,y_2)$ respectively. The $NS_3$ brane is extending in the $x^{4,5}$ directions and are placed at $x^{7,8,9}=0$, $x^6=y_{NS}$. In addition, there are $N$ suspended $D4$ branes between the $NS_3$ and $NS_2$ branes and a tilted $D4$ brane between the $NS_1$ and $NS_2$ branes.  See figure \ref{braneFig}.

By taking a field theory limit, in which the string length $l_s$ and brane distances go to zero but their ratios are fixed as shown below, we obtain the low-energy effective theory on the $N$ $D4$ branes. This effective field theory can be interpreted as a $U(N)$ gauge theory with 
vector-like quarks $q$, $\tilde{q}$ and a singlet chiral superfield $\Phi$ \cite{ISS}.
The superpotential is given by
\begin{equation}
W_{\rm  mag}=h  q_i{\Phi^i}_j \tilde{q}^j- h \mu^2 {\Phi^i}_i ,
\end{equation}
where $i,j =1,\cdots , N+1$ are flavor indices. According to Ref~\cite{Kutasov}, the parameters $h$, $\mu$ can be represented by the geometric data as
\begin{equation}
 h^2={8\pi^2g_sl_s\over \Delta y } , \qquad \mu^2={\Delta_4 \over 16\pi^3 g_s l_s^3} ,
\end{equation}
where $\Delta y=y_2 -y_1$. These parameters are fixed at finite values when we take the field theory limit. In the rest of the discussion, we will not assume the field theory limit although this limit is useful to understand the low-energy physics corresponding to the brane configuration. We keep the string length $l_s$ finite and take the string coupling small, which is called the brane limit \cite{IAS}.

Let us look at the false vacuum in this setup. At the tree level, the vacuum energy of the present configuration is $V_{\rm meta}=|h\mu^2|^2$. The $(N+1, N+1)$ component of ${\Phi^i}_j$, which we denote $X$, is massless and corresponds to the position of the tilted $D4$ brane in the $x^{7,8}$ directions. This brane can move freely in the $x^{7,8}$ directions which means that this is a flat direction of the brane configuration. However, there exist two kinds of corrections which induce a nonzero potential to the flat direction. One is the Coleman-Weinberg potential generated by an open string connecting between the tilted $D4$ brane and the horizontal $N$ $D4$ branes. The other comes from the gravitational effect of the $NS$-five branes. The explicit calculations of these effects have been done in Ref~\cite{Kutasov}. The resulting potential is
\begin{eqnarray}
V_{\rm correction}= C |h^2\mu|^2 |X|^2+\cdots ,\label{Bar}
\end{eqnarray}
where $C$ is a positive constant whose explicit form is given in Ref~\cite{Kutasov}.

\begin{figure}[!t]
\begin{center}
 \includegraphics[width=.32\linewidth]{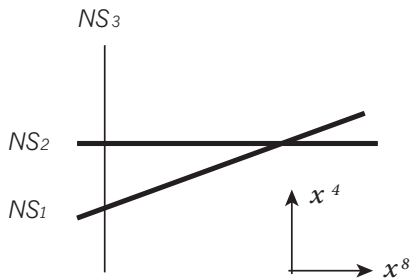}
\vspace{-.1cm}
\caption{The deformed brane configuration discussed in the text.}
\label{brane2Fig}
\end{center}
\end{figure}

We next consider a deformation of the present brane configuration without disturbing the metastability.
Rotate the $NS_1$ brane by a small angle $\varphi$ in the $(x^4,x^8)$ directions. See figure \ref{brane2Fig}. Then, there exists a configuration that the $D4$ brane suspended between $NS_1$ and $NS_2$ branes is parallel to the $N$ horizontal $D4$ branes and the length of the $D4$ brane is the shortest. This configuration corresponds to a SUSY preserving vacuum and is energetically favorable. In the field theory limit, the rotation can be interpreted as adding the mass term of $\Phi$ to the superpotential,
\begin{equation}
\Delta W={1\over 2} h^2 \mu_{\varphi}  \tr \Phi^2,\qquad \mu_{\varphi}={\tan \varphi \over 8\pi^2 g_s l_s},
\end{equation}
which allows us to solve the condition for a supersymmetric vacuum, $\partial W=0$. The supersymmetry preserving vacuum is placed at $X_{\rm SUSY}={\mu^2 / h\mu_{\varphi}}$. This new term also shifts the metastable vacuum slightly to the position $X_{\rm meta}={\mu_{\varphi}/ h C}$. It is convenient to introduce the quantities with dimension of length,
\begin{equation}
\Delta_8\equiv (X_{\rm SUSY}-X_{\rm meta})l_s^2,\qquad x_{\rm meta}\equiv X_{\rm meta}l_s^2.
\end{equation}
We will concentrate on this deformed brane configuration below.

The false vacuum finally decays into the supersymmetric vacuum by climbing up the potential barrier \eqref{Bar} and sliding down to $X_{\rm SUSY}$. 
The phase transition proceeds through a domain wall spreading out in the Minkowski space. This domain wall can be understood in terms of a $D4$ brane. The tilted $D4$ brane is located at $X_{\rm SUSY}$ in the true vacuum while $X_{\rm meta}$ in the false vacuum. Then, the domain wall corresponds to a $D4$ brane extending in the two dimensional internal space with boundaries given by these two configurations. The remaining two spacial directions of the $D4$ brane are in the Minkowski space and understood as the domain wall.

\subsection{A cosmic string/domain wall bound state}

\begin{figure}[!t]
\begin{center}
 \includegraphics[width=.26\linewidth]{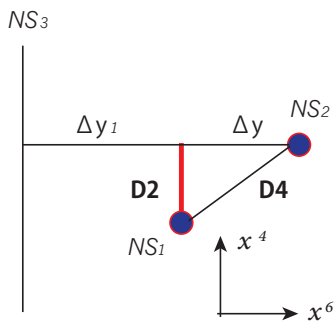}
\vspace{-.1cm}
\caption{A $D2$ brane corresponding to a cosmic string \cite{EHT}.}
\label{brane3Fig}
\end{center}
\end{figure}

We here introduce a cosmic string accommodated in the false vacuum. The authors of Ref~\cite{EHT} have discussed that a $D2$ brane extends from the $N$ $D4$ branes to the $NS_1$ brane in the $x^4$ direction and the remaining one spacial direction corresponds to a string in the Minkowski space. See figure \ref{brane3Fig}. In the present system, the phase transition between the false and true vacua proceeds through a domain wall which corresponds to a $D4$ brane as discussed above. In this case, the configuration that the $D2$ brane and the $D4$ brane exist separately is unstable. Instead, the $D2$ brane dissolves into the $D4$ brane and forms a bound state \cite{Text,Myers}. The dissolved $D2$ brane then yields a magnetic field on the $D4$ brane. The DBI action of the domain wall $D4$ brane is given as follows. We here assume that the domain wall is a tube-like object. Following the discussions in Ref~\cite{Emparan,Supertube,Hashimoto,hyakutake,PTK1,PTK2}, we consider the embedding function,
\begin{eqnarray}
&&X^0=t, \quad X^1=z, \quad X^2=R(t,z)\cos \theta ,\quad X^3=R(t,z)\sin \theta , \nonumber \\
&& X^4=x^4 \,\,\, \left( 0\le x^4 \le \Delta_4 \right),\quad X^8=x^8 \,\,\,  \left( x_{\rm meta} \le x^8 \le x_{\rm meta} +\Delta_8 \right), \label{embed}
\end{eqnarray}
where $X^{5,6,7,9}$ are constant. For simplicity, we further assume that the derivative of $R$ with respect to $z$ is zero, $R^{\prime}=0$. Since the cosmic string is along $z$ and $x^4$, the dissolved $D2$ brane gets the magnetic field in the $(\theta, x^8)$ directions. Thus, the low-energy effective action can be obtained by turning on $B\equiv 2\pi \alpha^{\prime}F_{\theta x^8}$. We have
\begin{equation}
\partial_{\alpha}X^{\mu}\partial_{\beta}X_{\mu}+2\pi \alpha^{\prime}F_{\alpha \beta}= \begin{pmatrix} 
 -1+\dot{R}^2 & 0 &0& 0& 0 \\ 
 0 & 1 &0 & 0 & 0 \\
   0 & 0 & R^2  & 0 & B \\
   0 & 0 &0 & 1 & 0 \\
     0 & 0 &-B  & 0 & 1
 \end{pmatrix}, \label{XXF}
\end{equation}
where $\alpha, \beta = t, z, \theta, x^4, x^8$ are the indices of the world-sheet coordinates of the $D4$ brane and $\mu, \nu=0,\cdots, 9$ are the space-time indices. $\dot{R}$ is the derivative of $R$ with respect to $t$. The (Euclidean) DBI action is then given by $(t \rightarrow i \tau)$ 
\begin{equation}
S_E= \int d \tau \left[ 2\pi T_{\rm DW}L  \sqrt{(1+\dot{R}^2)(R^2+B^2)}- \pi R^2 L \Delta V \right] .
\end{equation}
Here, $L$ is the length of the cosmic string and $T_{\rm DW}=T_{D4} \Delta_8 \Delta_4 I$ where $I$ is defined as 
\begin{eqnarray}
I&\equiv &\int_0^{1} du \, \sqrt{{\Delta y^2 \over \Delta_4^2}  +u^2} \nonumber \\
&=&\frac{1}{2} \left( \sqrt{ {\Delta y^2 \over \Delta_4^2}+1}+{\Delta y^2 \over \Delta_4^2} \log
   \left(\sqrt{{\Delta y^2 \over \Delta_4^2}+1}+1\right)- {\Delta y^2 \over \Delta_4^2} \log {\Delta y \over \Delta_4}\right) .
\end{eqnarray}
Note that $\Delta_8 \Delta_4 I$ is the area in $(x^4,x^6,x^8)$ filled by the domain wall $D4$ brane. $\Delta V$ is the difference of the energy densities of the two vacua. For later convenience, we have written the Euclidean action. The decay rate of this system at zero temperature has been studied in Ref~\cite{KO2}. In the next section, we would like to go a step further, including finite temperature effects. 

\section{Thermal potential}

Now, we are ready to study thermal effects on the inhomogeneous decay of the false vacuum.
Suppose that all the branes discussed in the previous section are in the thermal bath. The thermalized branes can be described by non-extremal black branes and affect dynamics of the tilted $D4$ brane via the gravitational effect. As we will see below, this effect can be identified with the thermal potential of the DBI action of the domain wall $D4$ brane.
At high temperature, the thermal effect drastically changes the shape of the potential and we have to consider the vacuum selection carefully
\cite{Kutasov}.
Instead, we assume sufficiently low temperature so that the energy of the supersymmetric vacuum is lower than that of our false vacuum.

 The most important effect on dynamics of the tilted $D4$ brane comes from the $NS_3$ brane.  The non-extremal five brane solution has been shown in Ref~\cite{Strominger}.
The metric at the location of the tilted $D4$ brane is given by
\begin{eqnarray}
&&d s^2=-f(r_{\rm NS})  d t^2 +   (d x^{i})^2 + (d x^4)^2 + (d x^5)^2 +H(r_{\rm NS}) \left[ f^{-1}(r_{\rm NS}) dr_{\rm NS}^2 +r^2_{\rm NS} d \Omega_{3}^2 \, \right], \nonumber \\[1ex]
&& e^{2(\phi- \phi_0)}=H(r_{\rm NS}),
\end{eqnarray}
where $i=0,\cdots , 3$, $\phi$ is the dilaton and
\begin{equation}
f(r_{\rm NS}) =1-{r_h^2 \over r^2_{\rm NS}},\qquad H(r_{\rm NS})=1+{l_s^2 \over r_{\rm NS}^2} .
\end{equation}
Here, $r_{\rm NS}$ is the distance from the $NS_3$ brane, $r^2_{\rm NS}\equiv \Delta y_1^2 +(x^7)^2+(x^8)^2+(x^9)^2$ where $\Delta y_1$ is the distance between the $NS_3$ and $NS_1$ branes in the $x^6$ direction, $\Delta y_1\equiv y_1-y_{NS}$. We assume $\Delta y \ll \Delta y_1$ and approximate the distance between $NS_3$ brane and the titled $D4$ brane in the $x^6$ direction by $\Delta y_1$.
Then, we use spherical coordinates $(r_{\rm NS}, \Omega_3)$ in the above metric. This non-extremal black brane gives an attractive force to the titled $D4$ brane. The thermal corrections from the other branes are sub-leading by the following reason: A $NS$-five brane has a tension proportional to $1/g_s^2$ while a $D4$ brane tension is proportional to $1/g_s$. Hence, for $g_s \ll 1$, the tension of a $NS$-five brane is much larger than that of a $D4$ brane and the effect on the gravitational background from a $NS$-five brane is more significant. In addition, the $NS_1$ and $NS_2$ branes are almost parallel for a small rotation angle $\varphi$, thus their attractive forces are canceled each other.

\begin{figure}[!t]
\begin{center}
 \includegraphics[width=.42\linewidth]{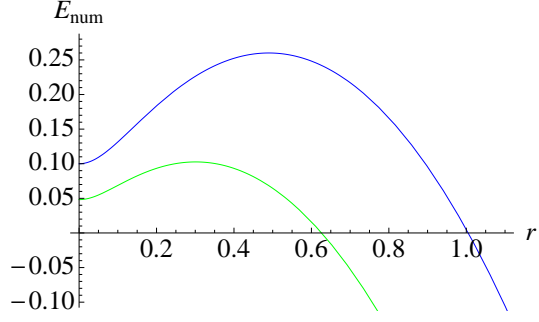}
  \vspace{0cm}
\caption{$E_{\rm num}$ as a function of $r$.
We take $\Delta Y=10$, $L_s=0$, $b=1/10$ and $\epsilon=1/10$. The blue and green curves correspond to the cases with $R_h=0,10$ respectively.}
\label{AAA}
\end{center}
\end{figure}

Let us consider the DBI action of the domain wall $D4$ brane with the magnetic flux from the dissolved $D2$ brane in the background of the non-extremal black brane.
We extend the zero temperature expression \eqref{XXF} to
\begin{equation}
\partial_{\alpha}X^{\mu}\partial_{\beta}X_{\mu}+2\pi \alpha^{\prime}F_{\alpha \beta}= \begin{pmatrix} 
 -f(r_{\rm NS})+\dot{R}^2 & 0 &0& 0& 0 \\ 
 0 & 1 &0 & 0 & 0 \\
   0 & 0 & R^2  & 0 & B \\
   0 & 0 &0 & 1 & 0 \\
     0 & 0 &-B  & 0 & H(r_{\rm NS})f^{-1}(r_{\rm NS})K(x^8)
 \end{pmatrix}.
\end{equation}
where $K$ is defined by
\begin{equation}
K(x^8)=1+(f-1){\Delta y_1^2 \over (x^8)^2+(\Delta y_1)^2}.
\end{equation}
The Euclidean DBI action is then given by 
\begin{equation}
S_E= \int d \tau \left[ 2\pi T_{\rm DW}L {1\over \Delta_8}\int d x^8 \sqrt{(f+\dot{R}^2) \left[Hf^{-1}K(x^8)R^2+B^2 \right]}- \pi R^2 L \Delta V \right] \label{Eaction}.
\end{equation}
We now define the following dimensionless quantities,
\begin{eqnarray}
&&{b}={\Delta V \over 2 T_{\rm DW}}B, \qquad r={ \Delta V \over 2 T_{\rm DW}}R,\qquad s={ \Delta V \over 2 T_{\rm DW}} \tau, \nonumber \\[1ex]
&&q\equiv  {x^8\over  x_{\rm meta}},\quad \epsilon\equiv {x_{\rm meta}\over \Delta_8}, \quad L_s\equiv {l_s \over  x_{\rm meta}},\quad  \Delta Y\equiv {\Delta y_1\over x_{\rm meta}},\quad R_h\equiv {r_h \over x_{\rm meta}}.\label{DimlessV}
\end{eqnarray}
By assuming a static solution, the energy of the string/domain wall bound state can be written in terms of these quantities as
\begin{eqnarray}
E&=&{4 \pi T_{\rm DW}^2 L \over \Delta V} \left[   \epsilon \int_{1}^{1+1/\epsilon} dq \sqrt{ \left(1+{L_s^2\over \Delta Y^2 +q^2 } \right)  K(q) \, r^2 +\left(1-{R_h^2\over \Delta Y^2 +q^2 } \right)  b^2}-r^2\right] \nonumber \\[1ex]
&\equiv & {4 \pi T_{\rm DW}^2 L \over \Delta V} \, E_{\rm num},
\end{eqnarray}
where we set $x^7 = x^9 = 0$ and $K(q)=1-{\Delta Y^2 R_h^2 \over (\Delta Y^2 +q^2)^2} $. Figure~\ref{AAA} shows the dimensionless energy ${E}_{\rm num}$ as a function of $r$. 
We take $\Delta Y=10$, $L_s=0$, $b=1/10$ and $\epsilon=1/10$. The blue and green curves correspond to the cases with $R_h=0,10$ respectively.
We can see that the hight and width of the energy barrier become small as $R_h$ is large and the thermal effect enhances the quantum tunneling probability that the domain wall expands to infinity. 

\begin{figure}[!t]
\begin{center}
 \includegraphics[width=.4\linewidth]{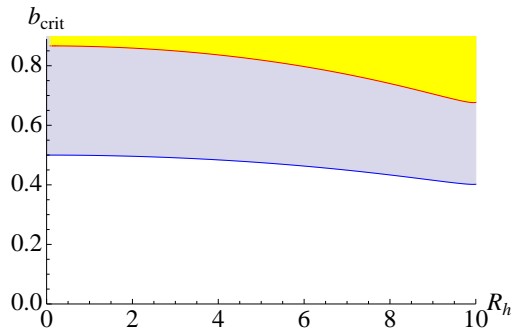}
\vspace{-.1cm}
\caption{A numerical plot of the condition that the domain wall always expands to infinity (the colored regions).
We take $\epsilon=1/10$ and $\Delta Y=10$.
In addition, $L_s=0, 10$ for the regions with the blue and red boundaries.}
\label{BBB}
\end{center}
\end{figure}

The domain wall expands with no bound when the value of $b$ reaches a critical point.
The condition that the domain wall always expands to infinity is given by
\begin{equation}
b \ge b_{\rm crit} \equiv \frac{\epsilon}{2} \int_1^{1+1/\epsilon} dq {H(r_{\rm NS})K(q) \over \sqrt{f(r_{\rm NS})}} \, . \label{stabilitycond}
\end{equation}
Figure \ref{BBB} shows 
a numerical plot of the condition that the domain wall always expands to infinity (the colored regions).
We take $\epsilon=1/10$ and $\Delta Y=10$.
In addition, $L_s=0, 10$ for the regions with the blue and red boundaries.

\section{Thermally assisted decay process}

In this section, we would like to go a step further toward the calculation of the decay rate of the false vacuum affected by thermally excited states. As we have mentioned in the introduction, since there are two types of thermal corrections, the complete analysis of the decay rate is quite involved. However,  fortunately, there is a parameter region where we can reliably neglect the thermal correction to the potential of the DBI action and focus on dynamics triggered by thermally excited modes. Roughly speaking, this parameter region corresponds to taking the distances between the branes large compared to the string length. We will discuss this region in detail at the end of this section.

\begin{figure}[!t]
\begin{center}
 \includegraphics[width=.4\linewidth]{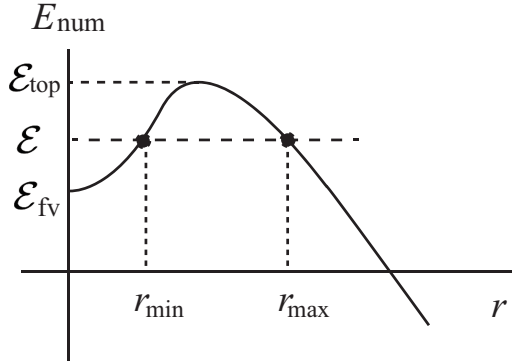}
\vspace{-.1cm}
\caption{The decay process by a thermally excited mode with the energy ${\cal E}$. We define the dimensionless quantities, ${\cal E}_{\rm top}\equiv {\Delta V \over 4\pi T_{\rm DW}^2 L}E_{\rm top}$ and ${\cal E}_{\rm fv}\equiv {\Delta V \over 4\pi T_{\rm DW}^2 L}E_{\rm fv}$.}
\label{energyFig}
\end{center}
\end{figure}

The thermal effect from excited modes can be described by imposing a periodic boundary condition on the Euclidean time in the domain wall $D4$ brane action. Let us ignore the thermal correction to the potential and write down the action with $f=1$ and $H=1$, 
\begin{equation}
S_E= \int d \tau \left[ 2\pi T_{\rm DW}L  \sqrt{(1+\dot{R}^2)(R^2+B^2)}- \pi R^2 L \Delta V \right] .
\end{equation}
Again, it is convenient to introduce the dimensionless quantities \eqref{DimlessV}. We can define the dimensionless action $S_{\rm num}$ as
\begin{eqnarray}
S_E&=&{4\pi T_{\rm DW}^2 L \over \Delta V   } \int ds \left[ \sqrt{(1+\dot{r}^2)(r^2+b^2)}- r^2 \right] \nonumber \\[1ex]
&=&{4\pi T_{\rm DW}^2 L \over \Delta V   } S_{\rm num} \, . \label{dimlessS}
\end{eqnarray}
The shape of the corresponding (dimensionless) energy is shown in figure \ref{energyFig}.
At finite temperature, the initial state of the phase transition can be not only the ground state but also thermally excited states. The thermal distribution of the excited modes with the energy $E$ is given by the Boltzmann distribution. The total tunneling rate is proportional to the following energy integral (for example, see Ref~\cite{E.Weinberg}):
\begin{eqnarray}
\Gamma &\propto& \int_{E_{\rm fv}}^{E_{\rm top}} dE \, e^{-\beta (E-E_{\rm fv})} e^{-B(E,T)} \nonumber \\[1ex]
&\simeq &  e^{-\beta (E_*-E_{\rm fv})} e^{-B(E_*,T)} , \label{totaltunneling}
\end{eqnarray}
where $\beta = 1/T$ and $B(E,T)$, a function of the energy and temperature $T$, is the bounce action which is given by the on-shell action subtracted by the static solution at $r = r_{\rm min}$
(see figure~\ref{energyFig})
\cite{Coleman}.
We will evaluate it below.
In addition, $E_{\rm fv}$ and $E_{\rm top}$ are the energies at $r=0$ and the top of the energy barrier respectively, as in figure \ref{energyFig}. In the second equality of the above equation, we have approximated the integral by the value of the integrand at its maximum.
Then, $E_*$ is the energy at the critical point of the exponent, which is obtained by the condition,
\begin{equation}
\beta=-{d \over d E} B(E,T)\Big|_{E = E_*}.\label{Periodic}
\end{equation}

Let us evaluate the bounce action to estimate the tunneling rate.
The equation of motion obtained from the dimensionless action $S_{\rm num}$ can be written in terms of the first order differential equation, 
\begin{equation}
{dr \over ds}=\pm {\sqrt{r^2+b^2-(\mathcal{E}+r^2)^2} \over (\mathcal{E}+r^2)} \, , \label{velocity}
\end{equation}
with the integration constant ${\cal E}$ corresponding to the dimensionless energy, which satisfies the relation,
\begin{equation}
\mathcal{E}=\sqrt{r_{\rm min}^2+b^2}-r_{\rm min}^2 \, . 
\end{equation}
At $r = r_{\rm min}$ and the bouncing point $r = r_{\rm max}$, the velocity \eqref{velocity} is zero and the following factorization condition is satisfied, 
\begin{equation}
r^2+b^2-(\mathcal{E}+r^2)^2=(r-r_{\rm min})(r_{\rm max}-r)(r^2+a_1r+a_0) ,
\end{equation}
where $a_0$ and $a_1$ are some constants.
This condition fixes $r_{\rm max}$ in terms of $r_{\rm min}$,
\begin{equation}
r_{\rm max}=\sqrt{1+r^2_{\rm min}-2\sqrt{b^2+r_{\rm min}^2}} \, .
\end{equation}
From the condition, $r_{\rm max}\ge r_{\rm min}$, we obtain the upper bound on $r_{\rm min}$, 
\begin{equation}
r_{\rm min}\le \sqrt{{1\over 4}-b^2} \, . \label{upperbound}
\end{equation}
Plugging the equation \eqref{velocity}  back into the dimensionless action, we obtain
\begin{eqnarray}
S_{\rm num}^{\rm on-shell}&=&\int ds \left[  {r^2+b^2 \over \mathcal{E}+r^2}-r^2 \right] \nonumber \\[1ex]
&=& \int_{r_{\rm min}}^{r_{\rm max}} dr {1\over \sqrt{r^2+b^2 -(\mathcal{E}+r^2)^2}}\left(r^2+b^2-r^2 ( \mathcal{E}+r^2) \right).\label{Bounce1}
\end{eqnarray}
In the second equality, we have changed the integral variable from $s$ to $r$ by using \eqref{velocity}.  On the other hand, the static solution of the action at $r=r_{\rm min}$ is given by
\begin{equation}
S_{\rm num}^{\rm sub}=   \int ds \, \mathcal{E} = \int_{r_{\rm min}}^{r_{\rm max}} dr {\mathcal{E}+r^2 \over \sqrt{r^2+b^2-(\mathcal{E}+r^2)^2}} \, \mathcal{E}. \label{Bounce2}
\end{equation}
Then, subtracting this solution from the action \eqref{Bounce1}, we obtain the (dimensionless) bounce action,
\begin{eqnarray}
\tilde{B} = S_{\rm num}^{\rm on-shell} - S_{\rm num}^{\rm sub} =\int_{r_{\rm min}}^{r_{\rm max}} dr \sqrt{r^2+b^2-(\mathcal{E}+r^2)^2} \, .
\label{bounceaction}
\end{eqnarray}
We use this bounce action to evaluate the total tunneling rate for the phase transition through the domain wall.

\begin{figure}[!t]
\begin{center}
 \includegraphics[width=.46\linewidth]{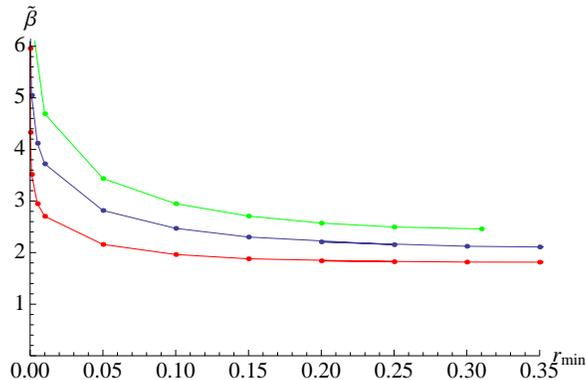}
\vspace{-.1cm}
\caption{The relationship between $\tilde{\beta}$ and $r_{\rm min}$. The green, blue and red polygonal lines correspond to the cases with $b=1/2.6,1/3, 1/4$ respectively. }
\label{RandBeta}
\end{center}
\end{figure}

We next find the explicit relation between (the inverse of ) temperature $\beta$ and $r_{\rm min}$
from the condition \eqref{Periodic}.
By using the expression of the dimensionless bounce action $\tilde{B}$, the condition can be  rewritten as
\begin{equation}
\tilde{\beta} \, \equiv \, {\Delta V \over 2T_{\rm DW}} \beta =\int_{r_{\rm min}}^{r_{\rm max}}{\mathcal{E}_*+r^2\over \sqrt{r^2+b^2-(\mathcal{E}_*+r^2)^2}} \, dr \label{period} \, ,
\end{equation}
where we have defined the dimensionless energy at the critical point, ${\cal E}_*\equiv {\Delta V \over 4\pi T_{\rm DW}^2L}E_*$.
Numerical estimations of this relation are shown in figure \ref{RandBeta}. The green, blue and red polygonal lines correspond to the cases with $b=1/2.6,1/3, 1/4$ respectively. From the figure, we can see that for $b=1/4$ there is a critical temperature at $\tilde{\beta}^{-1}\sim 1/2$ above which the upper bound \eqref{upperbound} is not satisfied.
In this region, the system is no longer stable and the thermal effect from excited modes destabilizes our cosmic string/domain wall bound state.

\begin{figure}[!t]
\begin{center}
 \includegraphics[width=.45\linewidth]{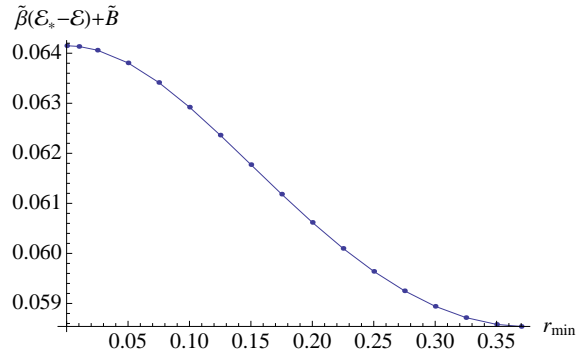}
\vspace{-.1cm}
\caption{The exponent of the decay rate monotonically decreases with the initial radius $r_{\rm min}$. We choose $b=1/3$. The thermally excited modes enhance the decay rate. }
\label{FigDecay}
\end{center}
\end{figure}

As we have obtained the expression of the bounce action \eqref{bounceaction} and understood the relation between $\tilde{\beta}$ and $r_{\rm min}$ \eqref{period}, we now estimate the total tunneling rate \eqref{totaltunneling}.
In this expression, the (dimensionless) thermally assisted factor is given by
\begin{equation}
\tilde{\beta} \left(\mathcal{E}_*(r_{\rm min})-\mathcal{E}_{\rm fv} (r=0) \right)=\tilde{\beta} \left(\sqrt{r_{\rm min}^2+b^2}-r_{\rm min}^2 -b \right).
\end{equation}
The bounce action \eqref{bounceaction}  is evaluated at $\mathcal{E} = \mathcal{E}_\ast$. 
Figure~\ref{FigDecay} shows a numerical plot for (the dimensionless part of) the exponent of the total tunneling rate. 
We can see that the exponent of the decay rate monotonically decreases as $r_{\rm min}$ is large. With the fact that $r_{\rm min}$ becomes large as the temperature $T$ increases (see figure \ref{RandBeta}), we conclude that the exponent of the total decay rate decreases at high temperature, which means that the thermal effect from excited modes enhances the tunneling rate and the life-time of the cosmic string/domain wall bound state becomes shorter.

Finally, we consider the validity of our approximation that the thermal correction to the potential of the DBI action, discussed in the previous section, is negligible. The correction to the potential is sub-leading when the temperature is lower than the string scale,
\begin{equation}
T l_s \ll 1 \label{AA} \, . 
\end{equation}
On the other hand, the thermal correction to the dimensionless bounce action presented in this section is roughly estimated as 
\begin{equation}
{2T_{\rm DW}\over \Delta V}T={\cal O} \left({\Delta_8 \over \Delta_4}\Delta y\, T \right) .
\end{equation}
Thus, when this quantity is larger than the gravitational corrections to the potential, our analysis is reliable. The condition is given by
\begin{equation}
{l_s \over \Delta y_1} \ll T{2T_{\rm DW}\over \Delta V}.
\end{equation}
Therefore, from this condition and \eqref{AA}, we find the parameter region where the thermal correction to the potential of the DBI action is negligible,
\begin{equation}
{l_s \over \Delta y_1} {\Delta_4\over \Delta_8  }{1\over \Delta y}\ll T\ll {1\over l_s} \, .
\end{equation}
The parameter space corresponds to taking the distances between the branes large compared to the string length.

\section{Conclusions and discussions}

In this short paper, we have investigated thermal effects to the decay rate of a false vacuum realized in Type IIA string theory. The false and true vacua are simultaneously realized in a single brane setup. We have discussed an inhomogeneous vacuum decay triggered by a cosmic string. The string and the domain wall are described by $D2$ and $D4$ branes, and because of the instability of the $D2/D4$ brane system, they form a bound state. That is, the $D2$ brane dissolves and a nonzero magnetic field is induced on the $D4$ brane. We have discussed two types of thermal effects when the brane system is put into the thermal bath. One is the correction to the potential of the DBI action and the other is the decay of the false vacuum via thermally excited states. To read off the underlying physics of these two effects, we have taken a region in the parameter space where one of the corrections is negligible. For the thermal correction to the potential, we have shown that the hight and width of the energy barrier become small as $R_{h}$ is large, while thermally excited modes tend to destabilize the domain wall/string bound state. Hence, in total, we conclude that the thermal effects destabilize the domain wall/string bound state and make the life-time of the false vacuum shorter.

For future directions, it would be interesting to extend our analysis to the model with the compact internal space such as the KKLT model \cite{KKLT}. In addition, uplifting our study to M-theory is an interesting avenue to show generality of our idea. Metastable vacua have been constructed in the perturbed Seiberg-Witten theory \cite{SWoriginal} and also the geometrically engineered theories
\cite{GeoMeta,SWMeta}.
These vacua may give us good toy models to discuss the decay process of the false vacua in M-theory. These topics are beyond the scope of this paper and we would like to revisit in future studies.

\section*{Acknowledgement}

YN is supported by a JSPS Fellowship for Research Abroad.
YO is supported by Grant-in-Aid for Scientific Research from the Ministry of Education, Culture, Sports, Science and Technology, Japan (No. 25800144 and No. 25105011).

%
%

\end{document}